\documentstyle[epsf]{EuroPhys}
\input EuroMacr

\begin{document}

\newcommand{\DS}{\displaystyle}
\newcommand{\SS}{\scriptstyle}
\newcommand{\SSS}{\scriptscriptstyle}
\newcommand{\lsim}{\stackrel{\SSS <}{\SSS \sim}}
\newcommand{\gsim}{\stackrel{\SSS >}{\SSS \sim}}

\euro{}{}{}{}
\Date{}

\title{The role of long-range forces in the phase behavior of colloids and proteins.}
\shorttitle{}
\author{M. G. Noro, N. Kern \And D. Frenkel }
\institute{
      FOM Institute for Atomic and Molecular Physics, 
      Kruislaan 407, 1098 SJ Amsterdam, The Netherlands
          }
\rec{}{}
\pacs{
\Pacs{82}{70Dd}{Colloids}
\Pacs{87}{15Da}{Physical chemistry of solutions of biomolecules; 
		condensed states}
\Pacs{64}{75}{Solubility, segregation and mixing}
}

\maketitle

\abstract{\ The phase behavior of colloid-polymer mixtures, and of solutions
of globular proteins, is often interpreted in terms of a simple model of 
hard spheres with short-ranged attraction. While such a model yields a 
qualitative understanding of the
generic phase diagrams of both colloids and proteins, it fails to capture
one important difference:  the model predicts fluid-fluid phase separation 
in the metastable regime below the freezing curve. Such demixing  has been 
observed for globular proteins, but for colloids it appears to be pre-empted 
by the appearance of a gel. In this paper, we study the effect of additional 
long-range attractions on the phase behavior of spheres with 
short-ranged attraction. 
We find that such attractions can shift the (metastable) fluid-fluid
critical point out of the gel region. As this metastable
critical point may be important for crystal nucleation, our results suggest
that long-ranged attractive forces may play an important role in  the
crystallization of globular proteins. However, in colloids, where refractive
index matching is often used to switch off long-ranged dispersion forces,
gelation is likely to inhibit phase separation.}

\section{Introduction}
%
X-ray crystallography is still the standard technique to resolve the
three-dimensional structure of globular proteins. But crystallography
requires crystals, and protein solutions are notoriously difficult to
crystallize. In order to understand the factors that favor crystallization,
it is useful to gain insight into the phase behavior of the protein solution.
As a first approximation, it is often sufficient to consider proteins as hard
spherical bodies, interacting through a short-ranged attractive potential.
In fact, Rosenbaum et al. \cite{RZZ} have shown that the
crystallization curves for a number of globular protein solutions appear to
coincide with those of a system of hard spheres with a rather short-ranged
attractive Yukawa interaction, as studied in the
simulations of Hagen and Frenkel \cite{Hagen}. These simulations were primarily aimed at modeling the phase behavior of polymer-colloid mixtures. 
\\
As the attractive Yukawa model is used to model both colloids and globular
proteins, one should expect that the conclusions that hold for one system
should be transferable to the other. This does indeed appear to be correct,
as far as the equilibrium phase behavior is concerned: experimental studies
of colloid-polymer mixtures \cite{Ilett} show that, as the range of the
attractive interaction between the colloids is shortened, the phase diagram
changes in the way predicted originally by Gast et al. \cite{GRH} and
subsequently these predictions were analysed in considerable detail, both by computer simulation 
\cite{Meijer} and theoretically \cite{HNWL}. In particular, this analysis
shows that fluid-fluid coexistence occurs only if the range of the attraction
is sufficiently large compared to the ``hard-core'' radius of
the particle (typically, more than 30\%). For shorter-ranged attractions,
the stable fluid-fluid transition is pre-empted by freezing. There are several 
experimental studies that indicate that solutions of globular proteins may 
exhibit the phase behavior expected for spherical particles with short-ranged 
attraction~\cite{Berland}.
In the context of protein crystallization, the presence of a metastable
fluid-fluid coexistence curve and, in particular, of a metastable critical
point may be important, as ten Wolde and Frenkel \cite{PR} have argued that the
presence of such a metastable critical point will lower the barrier for
crystal nucleation. 
\\
However, experimental studies of suspensions of colloids with a short-ranged
attractive interaction suggest that there is an important difference in the
phase behavior of proteins and colloids: whereas a metastable
fluid-fluid coexistence curve has actually been observed for several
globular proteins, colloids with short-ranged attractive interactions tend
to form a gel-like phase instead. Although the latter phase is metastable, 
it can delay \cite{Smits90}, or even suppress, crystallization \cite{Ilett}. 
Clearly, the model of  (mono-disperse) hard spheres with short-ranged
attractive interactions is an oversimplification.
Real proteins are non-spherical and have non-isotropic (``patchy'')
interactions. In contrast, while colloids may be quite spherical, they are
hardly ever monodisperse. All these factors will affect the tendency to
crystallize, to phase-separate and to form a gel. 
In the present paper, we focus
on a very simple phenomenon, namely the effect of long-range forces. 
The reason why we focus specifically on long-range forces 
(rather than on poly-dispersity or particle anisometry) is that we are looking for
 a mechanism that can move the gelation regime well below the fluid-fluid critical
 point. We shall argue that long-range attractive interactions do precisely that. 

\section{Model and equations of state}
%
We wish to consider a system that can exhibit freezing, fluid-fluid phase 
separation and gelation. 
In our model two spherical hard particles of diameter 
$\sigma$ experience a short-ranged attraction through a potential of the form:
\begin{equation}
  u(r) = 
    \left\{ 
     { \matrix
         {\infty \hfill \cr 
           -\epsilon \cdot (r/\sigma)^{-n} \hfill \cr 
         }
     }\right. 
  \quad 
  \matrix
  {
    {r\le \sigma }\hfill \cr 
    {\sigma <r}\hfill \cr 
  }  	\qquad.
  \label{eq1}
\end{equation}
The specific form of the interaction potential has been chosen for convenience.
In the case of proteins, we do not really know the detailed form of the attractive 
interaction (other than that it has a short-ranged component). 
In mixtures of ``hard-core" colloids and non-adsorbing polymer, the short-ranged
 attraction is induced by depletion forces (see \cite{HNWL} and references therein). 
Other functional representations could have been chosen for the short-ranged 
potential (and have indeed been considered in studies of fluid-fluid 
coexistence and percolation \cite{Xu}). However, the 
choice should become unimportant for very short-ranged attractions 
\cite{Regnaut}. The fluid is then 
well described by Baxter's adhesive hard-sphere model of infinitely short-ranged 
attraction \cite{Baxter}. 
An approximate
equation-of-state for this model is known (see the first two terms in eq. \ref{Eq-FluidPressure} below, given 
in ~\cite{Baxter,Barboy}). The only parameter is the value of the 
second virial coefficient, usually expressed in terms of a ``stickiness'' 
parameter $1/\tau^{SS}$:
\begin{equation}
  B_{2}^{SS}
    \equiv B_2^{HS} \ (1-\frac{1}{4\ \tau ^{SS}})
  \qquad,
\end{equation}
where $B_{2}^{HS}=2\pi \sigma ^{3}/3$ is the virial coefficient of hard spheres.
$\tau ^{SS}$ can be thought of as a measure of the temperature. In particular, 
the limit $\tau ^{SS}\rightarrow \infty $ corresponds to the situation where 
the effect of short-ranged attraction becomes negligible. 
\\

In the present case, the second virial coefficient is given by: 
\begin{equation}
  B_2^{HS}\ (1-\frac{1}{4\ \tau ^{SS}(t)})
  =
  B_2^{HS}
    - 2 \pi \int_{\sigma}^{\infty} 
            \left[ e^{1/t \cdot (r/\sigma)^{-n}} - 1 \right] 
          \ r^2 dr
  \qquad.
  \label{Eq-VirialCoeff}
\end{equation} 
$t \equiv k_BT/\epsilon$ is the dimensionless temperature.
In the present approach we account for the additional
long-range attraction, by adding
a Van-der-Waals like contribution to the equation of state: 
\begin{equation}
  \frac{p_{fl}\ v_{0}}{k_{B}T}
  =   \eta \cdot \frac{1+\eta +\eta ^{2}}{(1-\eta)^{3}}
    - \eta ^{2}\cdot \lambda \frac{18 (2+\eta )
    - \lambda ^{2}\eta }{36 (1-\eta)^{3}}
    - \frac{\alpha_0}{t} \cdot \eta ^{2} 
  \qquad.
  \label{Eq-FluidPressure}
\end{equation}
p is the pressure, $v_{0}$ the hard sphere volume, $\eta$ the volume
fraction. $\alpha_0$ is related to the usual Van-der-Waals parameter $a$
by $\alpha_0 = a/(\epsilon v_0)$, and thus measures the long-range
attraction in units of $\epsilon$. The stickiness enters through the parameter 
$\lambda =\lambda (\tau ^{SS},\eta )$, given in ref. \cite{Baxter}.
The first term on the right-hand side of eq. (\ref{Eq-FluidPressure}) describes 
the hard-sphere contribution to the pressure, the second term accounts for 
the stickiness, and the third term describes the effect of long-ranged 
attraction.


To describe the solid phase, we follow Daanoun et al. \cite{Tejero},
estimating its entropy by a cell-theory and its energy by
the mean field interaction energy between nearest neighbors at their 
average positions. The equation-of-state of the solid is most readily 
expressed in terms of a re-scaled distance,
$s\equiv (\eta_{cp}/\eta )^{1/3}$, where $\eta _{cp}$ is the volume 
fraction at regular close packing: 
\begin{equation}
  \frac{p_{sol}\cdot v_{0}}{k_{B}T}
    =  \eta _{cp}\cdot \left[ \frac{1}{s^{2}(s-1)}
     + \frac{1}{t}\frac{z \ u ^{\prime}(s)}{6s^{2}}\right] 
     - \frac{\alpha_0}{t} \cdot \eta ^{2} 
  \qquad. 
\end{equation}
$z$ denotes the number of nearest neighbors (here $z=12$ as for the FCC 
lattice). 

\section{Gelation}
%
In the gel state the colloids form a space-spanning network. 
The tendency to form gels depends strongly on the range and
strength of the forces acting between the particles. 
There appears to be no general recipe to predict whether
a given system should be in the gel state. However, in the case of adhesive
spheres, there is a simple analytical expression~\cite{CG}, that allows us to estimate the percolation curve. In a system of adhesive spheres, percolation is a necessary, but not
a sufficient condition to form a gel. Hence, we can use the percolation
criterion of ref.~\cite{CG} to delimit the region where gelation
is possible. For purely adhesive spheres, the analysis of ref. \cite{CG}
indicates that cluster percolation occurs if 
\begin{equation}
  \tau ^{SS}\le \frac{1-2\eta +19\eta ^{2}}{12\ (1-\eta )^{2}}
  \qquad.
  \label{Eq-PercolationLine}
\end{equation}
One might expect that a different percolation criterion should apply if the
particles have a longer-ranged attraction, in addition to the ``sticky''
interactions. However, long-ranged attractions should have
little effect on the percolation. In fact, in the true van-der-Waals limit
(infinitely weak, infinitely long-ranged attractions) the long-ranged
attractions do not affect the structure of the fluid at all, and have
therefore no effect on the percolation transition. It seems plausible that
attractions with a long, but finite, range will have a small effect
on percolation. Indeed, Kaneko \cite{Kaneko} has shown that
this is even true for electrostatic long-range interactions. In what
follows, we shall therefore continue to use eq. (\ref{Eq-PercolationLine})
to delimit the regions where no percolation occurs, and where gelation can
thus safely be ruled out.

\section{Results}
%
Using the equations of state given above, we study the effect of
additional long-range attraction on the phase diagram of particles with a
short-ranged attraction. We locate the phase-coexistence
boundaries numerically, by imposing that the pressures and the chemical
potentials of the coexisting phases be equal.
In order to evaluate the effect due to the long-range attraction, we have to
compare to the case of pure sticky spheres. In analogy with the stickiness
parameter $\tau ^{SS}$, we define a second parameter $\tau ^{VdW}$, such
that $\tau ^{SS}$ and $\tau ^{VdW}$ characterize long- and short-range
interactions separately: 
\begin{equation}
  B_{2} = B_{2}^{HS}\ 
    \left( 
      1-\frac{1}{4\ \tau ^{SS}}-\frac{1}{4\ \tau ^{VdW}}
    \right)
  \qquad.
\end{equation}
Both parameters depend on temperature. Whereas we have simply
$\tau^{VdW} = t/\alpha_0$ from eq. (\ref{Eq-FluidPressure}), $\tau^{SS}$ 
is obtained (numerically) from eq.(\ref{Eq-VirialCoeff}) for $n=50$. 
To compare the phase diagrams of different model systems, it is convenient to introduce the {\it effective} stickiness, defined as 
\begin{equation}
  \frac{1}{\tau ^{eff}}\equiv \frac{1}{\tau ^{SS}}+\frac{1}{\tau ^{VdW}}
  \qquad.
  \label{Eq-tauEff}
\end{equation}
In what follows, we shall compare phase diagrams as a function of ${\tau^{eff}}$.

%
%
%
\begin{figure}[tbp]
\hfil
\begin{minipage}{4.7cm}
    \epsfxsize 4.0 cm
    \epsfbox{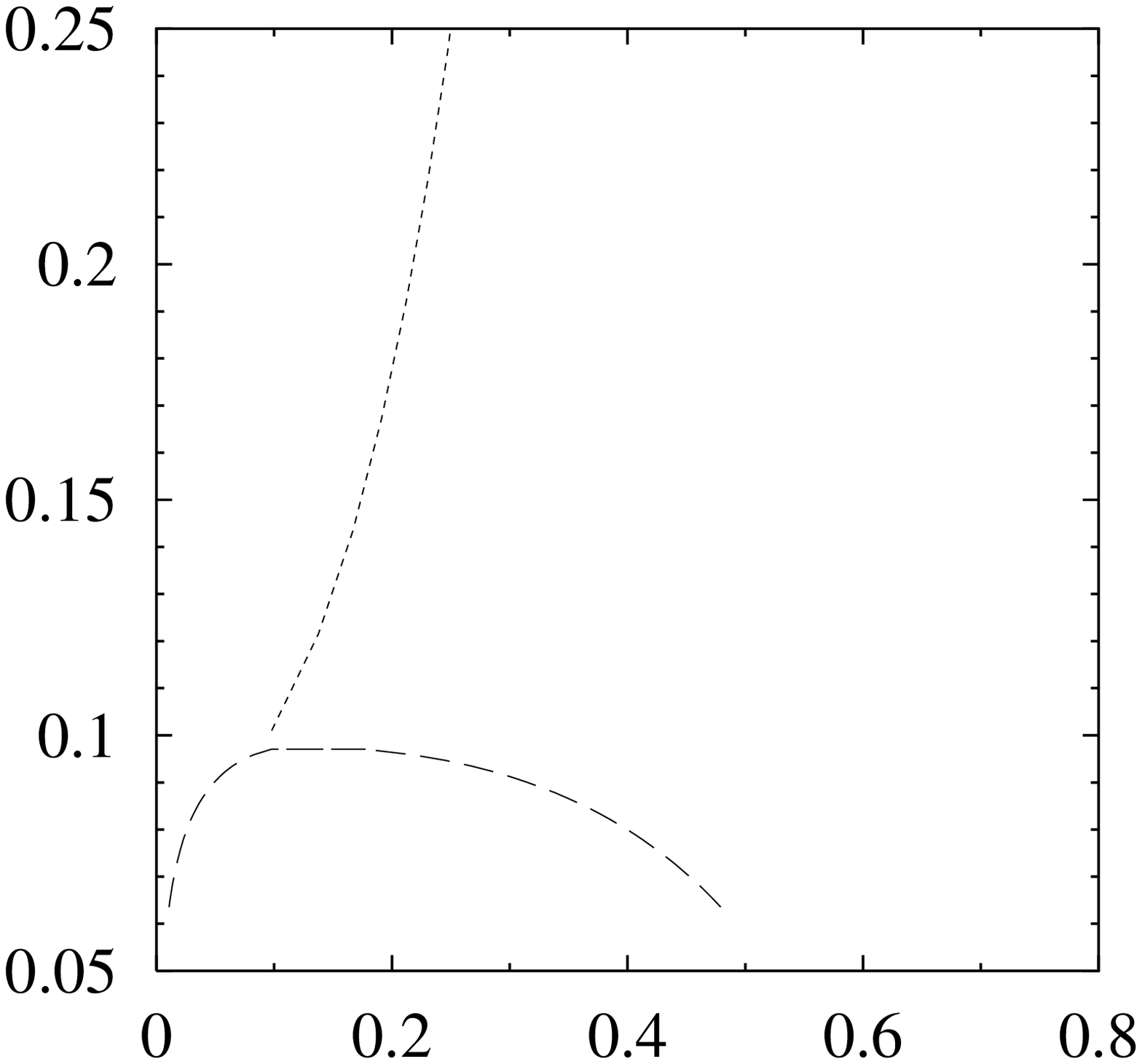}
    \center{(a) $\alpha_0=0.0$}
  \end{minipage}
\setlength{\unitlength}{1cm} \put(-5.1,1.2){$\tau^{eff}$} \put(-0.5,-2.0){$%
\eta$} 
\put(-3.2,0.0) 
{ \begin{minipage}{2.7cm} 
    \epsfxsize 2.0 cm 
    \epsfbox{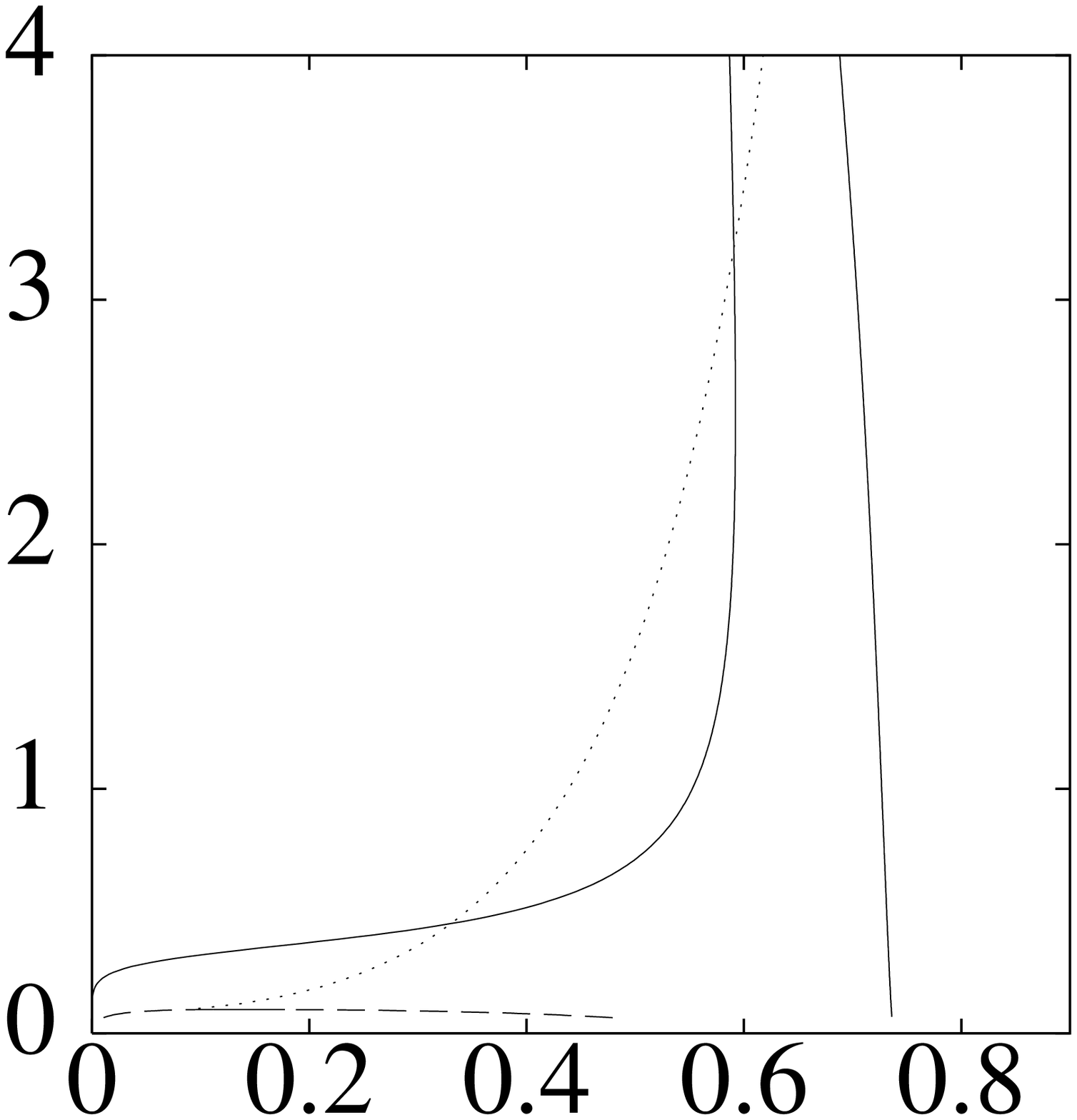} 
  \end{minipage} } 
\hfil
\begin{minipage}{4.7cm}
    \epsfxsize 4.0 cm
    \epsfbox{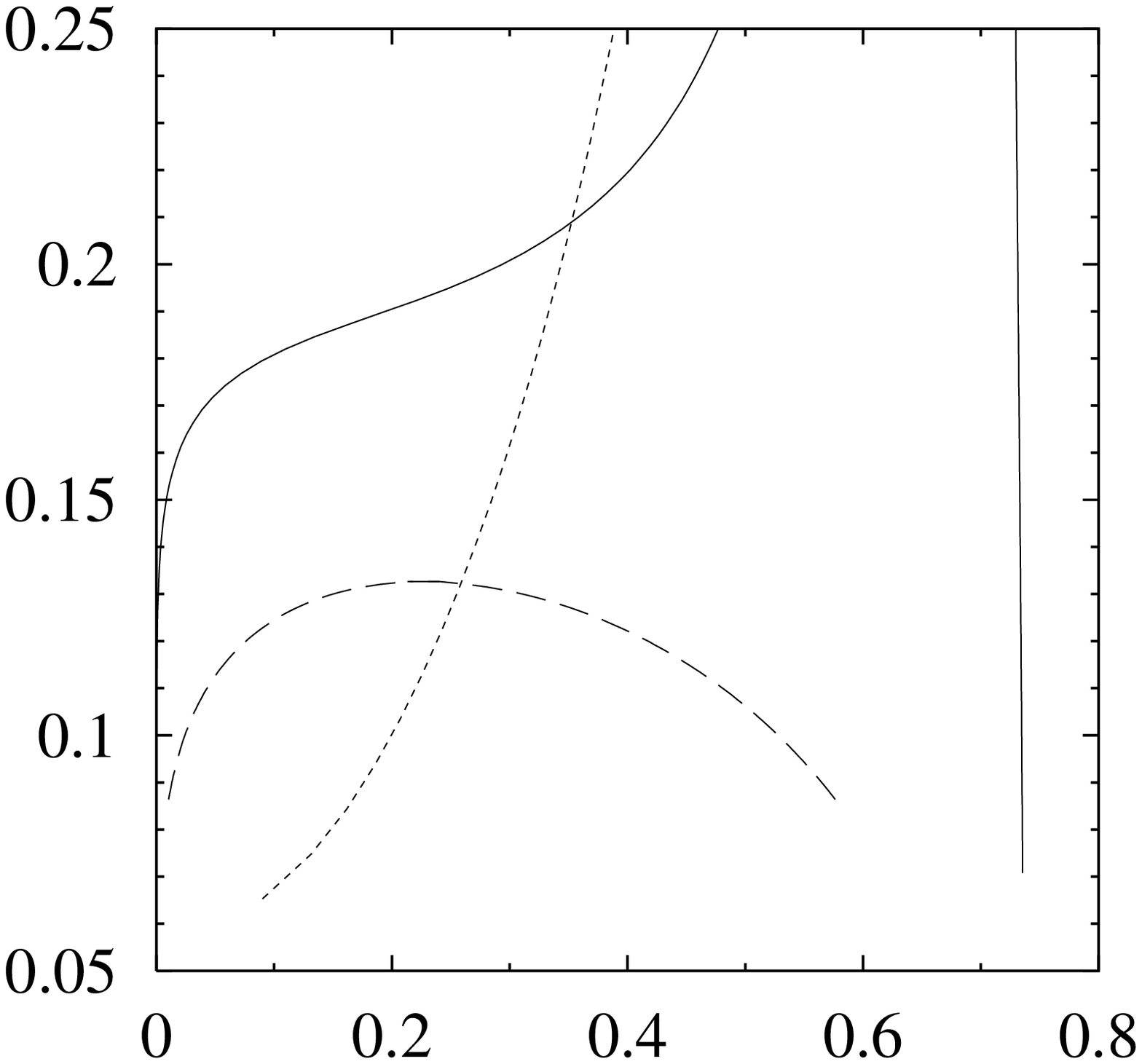} 
    \center{(b) $\alpha_0=1.0$}
  \end{minipage}
\setlength{\unitlength}{1cm} \put(-5.1,1.2){$\tau^{eff}$} \put(-0.5,-2.0){$%
\eta$} \hfil
\begin{minipage}{4.7 cm}
    \epsfxsize 4.0 cm
    \epsfbox{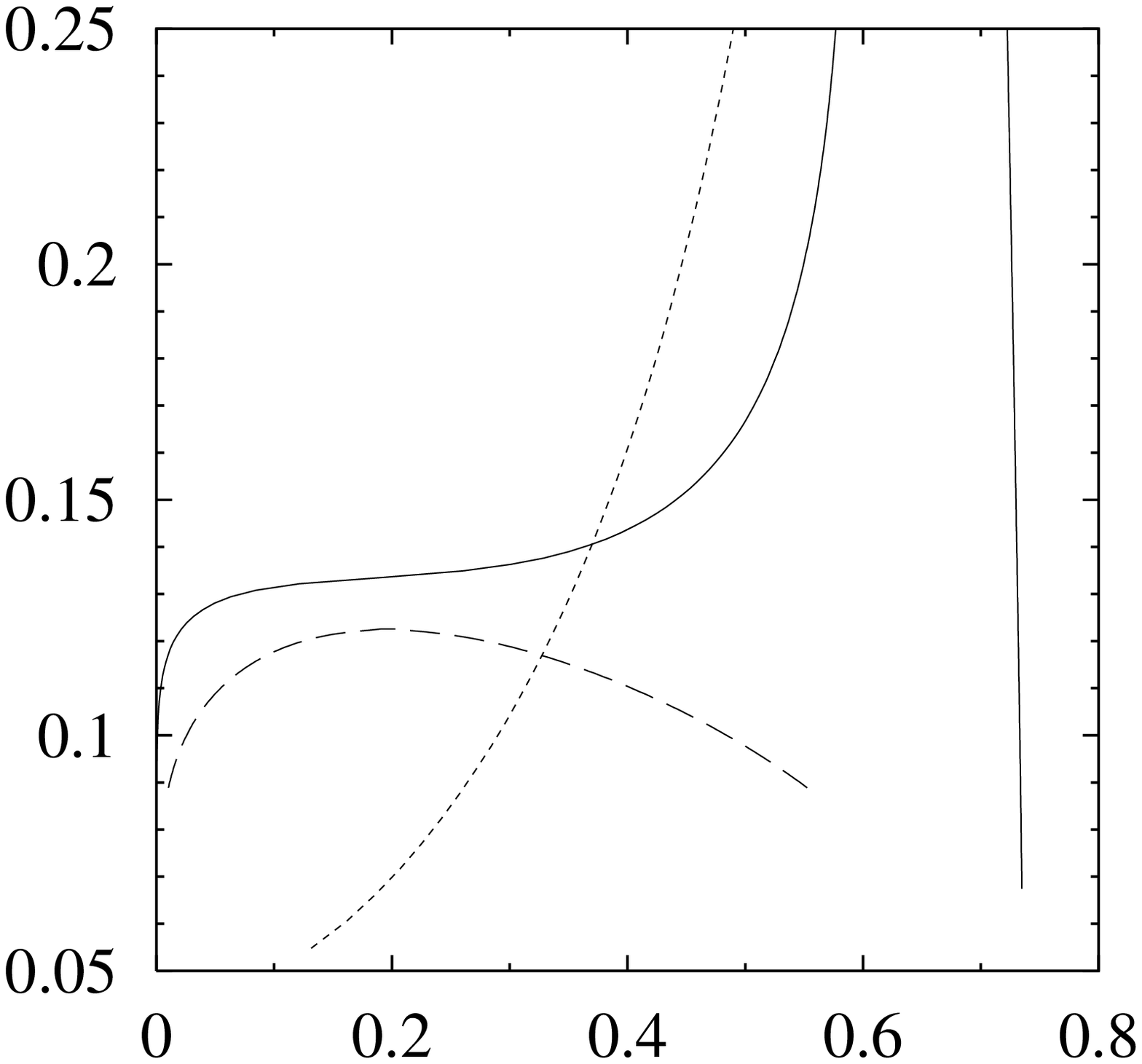} 
    \center{(c) $\alpha_0=2.0$}
  \end{minipage}
\setlength{\unitlength}{1cm} \put(-5.1,1.2){$\tau^{eff}$} \put(-0.5,-2.0){$%
\eta$}
\caption{Phase diagrams in the $(\tau ^{eff},\eta )$ representation, shown
for increasing Van-der-Waals attraction. Every plot shows the fluid-fluid
coexistence (broken line), the fluid-solid coexistence (solid line) as well 
as the percolation line (dotted line). As the long-range attraction
becomes more important, the fluid-fluid critical point shifts out of the
region where percolation can occur. For the solid, $n=50$ has been used. }
\label{Fig-PhaseDiagrams}
\end{figure}
%
%
The main results of our analysis are presented in fig. (\ref{Fig-PhaseDiagrams}).
Without the Van-der-Waals contribution, fig. (\ref{Fig-PhaseDiagrams}.a),
$\tau^{eff}$ is identical to $\tau ^{SS}$, and we simply obtain the phase 
diagram of spheres with purely short-ranged attraction. 
The fluid-fluid critical point is
metastable and lies {\it below} the percolation line. This implies that
percolation, and presumably gelation, occurs before the system can be
quenched to the metastable critical point. Hence, in such systems one
should not expect to observe a fluid-fluid phase separation.
\\
In fig. (\ref{Fig-PhaseDiagrams}.b), a moderate Van-der-Waals
attraction is present. Due
to the long-range attraction, the percolation line now occurs for lower values of $\tau ^{eff}$, as it only depends on $\tau^{SS}$. In
contrast, the fluid-fluid coexistence is not strongly affected by the
different range. It consequently shifts out of the percolation region, and
it becomes possible to quench to this point without risk of gelation.
\\
As the long-range attraction is increased further,
fig. (\ref{Fig-PhaseDiagrams}.c), the same trends are observed~: 
the fluid-solid line
approaches the fluid-fluid critical point, which remains clearly outside the
percolation region. For even stronger long-range attraction, the
fluid-fluid critical point becomes stable. This reflects the fact that a
sufficiently strong Van-der-Waals attraction can, of course, force the fluid
to phase separate, whether short-range attractions are present or not.
%
%
%
%
\begin{figure}[tbp]
  \begin{minipage}{4.5cm}
    \epsfxsize 6.0 cm
    \epsfbox{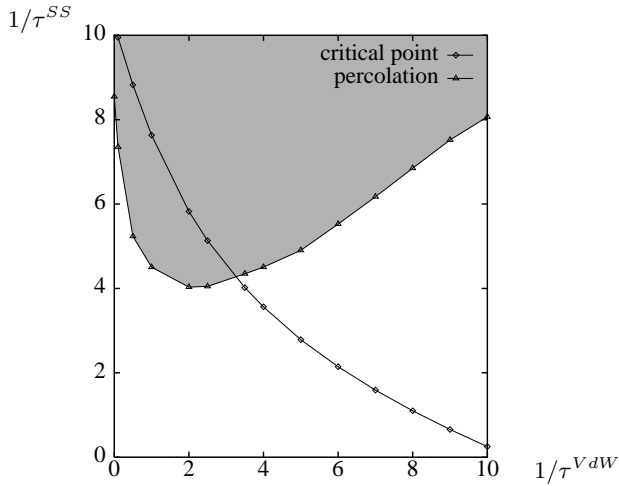} 
  \end{minipage}
    \setlength{\unitlength}{1cm}
     \put(-6.5,3.0){$1/\tau^{SS}$}
     \put(0.5,-3.0){$1/\tau^{VdW}$}
\par
    \caption{
    \label{Fig-tauDiagram}
    The curve connecting the diamonds delimits the region
    (in the upper right-hand corner of  ($1/\tau^{SS}$,$1/\tau^{VdW}$)-plane) 
    where fluid-fluid coexistence can occur. The curve thus represents the 
    collection of the critical points. The curve connecting the triangles 
    delimits the region (shaded) where percolation is expected to occur. 
    The curve was constructed by plotting the value of the percolation 
    threshold evaluated at the fluid-fluid critical density, at fixed $\tau^{VdW}$.
    In absence of long-range 
    attraction, percolation precedes fluid-fluid demixing. However, even 
    a moderate long-ranged attraction moves the fluid-fluid coexistence 
    curve out of the gelation region. 
    }
\end{figure}
%
\\
This result is summarized in fig. (\ref{Fig-tauDiagram}).
To read this figure, first consider the $1/\tau ^{VdW}$-axes, corresponding to 
long-ranged attraction only: when the attraction exceeds a certain value, 
fluid-fluid demixing occurs. Similarly, for  purely sticky spheres 
demixing occurs when 
$1/\tau ^{SS}\stackrel{\scriptscriptstyle >}{\scriptscriptstyle \sim } 10.9$,
When both types of interaction are present, the critical point
deviates a little from a line of constant $1/\tau^{eff}$.
Percolation on the other hand can only occur in the shaded area. 
The figure shows that, as the strength of the long-range attraction increases 
(increasing $1/\tau ^{VdW}$), the fluid-fluid critical point moves to  
a region where no gel formation is possible.

\section{Discussion}
It is tempting to speculate that the difference in phase behavior of
globular proteins and colloids may, at least partly, be due to the
different role of long-ranged attractive forces. In most studies of the
phase behavior of colloid-polymer mixtures, the refractive index of the
solvent is matched to that of the colloidal particles to facilitate
light scattering or microscopy studies of the colloidal structure. But, by
refractive-index matching, the attractive dispersion forces between the
colloids are effectively switched off. Hence, such suspensions are expected to 
behave as the model system with purely short-ranged
attraction, for which the metastable fluid-fluid
critical point lies well within the percolation region and fluid-fluid 
demixing is pre-empted. This is the state of affairs observed in the
experiments of ref. \cite{Ilett}. In contrast, recent experiments by Hachisu~\cite{Hachisu} on colloids with strong  van der Waals interactions show that, in such systems, fluid-fluid demixing does occur. 
Protein-protein interactions have been extensively studied both theoretically
and experimentally\cite{Lenhoff_exp,Lenhoff_the}.  However even the most careful
theoretical calculations, accounting for steric, electrostatic, dispersional
and short-range interactions, yield estimates for the second virial 
coefficient that may differ appreciably from the experimental values. 
In other words, much less is known about the  long-ranged  attractions between globular proteins (not necessarily dispersion forces) than those acting between  colloids. 
As we have shown above, the presence of
moderately strong long-ranged attractions would move the fluid-fluid
critical point outside the regime where gelation is likely to occur. In such
systems, the metastable fluid-fluid phase separation may be observed, as
indeed it is in the experiments of ref. \cite{Berland}.
Of course, there are many other factors that play a role in phase behavior
of proteins and colloids. For instance, colloids are usually slightly
poly-disperse, while proteins are not. It seems likely that the main effect
of poly-dispersity will be to bring down the fluid-solid coexistence curve
to lower values of $\tau $. However, the effect on the fluid-fluid phase
transition and on the percolation curve is expected to be small. The same is
expected for the effect of increasing the range of the ``short-ranged''
attraction. In contrast, anisotropic interactions (be they due to
non-sphericity or surface ``patchiness'') are likely to increase the
tendency to form gels. Clearly, the role of long-ranged attractions is only
one out of many - but at least it is an effect that provides a simple
explanation why metastable fluid-fluid separation is observed in globular
proteins but not in colloid-polymer mixtures.
\\

Let us finally consider the role of the metastable spinodal in crystal
nucleation. In the model studied in ref. \cite{PR}, crystal nucleation was
facilitated by the vicinity of the metastable critical point. Moreover, in
this model system, gelation did not interfere with either fluid-fluid phase
separation or crystal nucleation. However, as fig. (\ref{Fig-PhaseDiagrams})
shows, we should expect that in many cases gelation interferes with
fluid-fluid phase separation - if not in the early stages, than at least in
the later stages. Poon \cite{Poon} has argued that it is the metastable
fluid-fluid phase separation {\it itself} that leads to gelation, and
thereby inhibition of crystallization. In the case of colloid-polymer
mixtures,  phase separation and gelation tend to occur in the same region of
the phase diagram. However, we stress that it is important to identify
gelation as the primary phenomenon that suppresses crystallization -
fluid-fluid phase separation may also be suppressed by gelation, but it does
not {\it cause} gelation. In fact, the simulations of ref. \cite{PR} provide
a nice example of fluid-fluid phase separation without gelation: in that
case the metastable fluid-fluid critical point actually {\em enhances }%
crystal nucleation. 
The results of ref. \cite{PR} suggest that, above the metastable
fluid-fluid coexistence curve, the early stages of crystal nucleation
involve the formation of a liquid-like nucleus. If this happens, gelation
can still interfere with the crystallization process at a later stage,
namely through the gelation of the liquid-like pre-critical (and, below the fluid-fluid binodal, even post-critical) nuclei. This
would slow down the subsequent formation of crystals. However, it is
unlikely that gelation would completely immobilize the particles in
small clusters - in fact, recent experiments on vitrification in thin films
indicate that the effective glass-transition temperature may be lowered
appreciably as the linear dimensions of the system are reduced \cite{Jerome}.
\\
Finally, we note that Sear \cite{RichardSear} has pointed out one more reason
why systems with short-ranged attractive interactions will not crystallize
easily: in the relevant part of the phase diagram, the interfacial
free energy $\gamma _{solid-fluid}$ of such systems tends to be large
compared to the thermal energy. According to classical nucleation theory, 
the crystal-nucleation barrier is proportional to the cube power of $\gamma
_{solid-fluid}$. Hence deeper quenches are needed to  get appreciable
nucleation rates. But, of course, the deeper the quench, the more likely it
is that gelation will interfere. However, Sear's argument does not distinguish between colloids and proteins.

\stars
This work was supported in part by `Chemische Wetenschappen-Unilever' program  with financial aid from both NWO (`Nederlandse Organisatie voor Wetenschappelijk
Onderzoek') and Unilever Research. 
The work of the FOM Institute is part of the research program
of ``Stichting Fundamenteel Onderzoek der Materie'' (FOM) and is supported
by NWO. We gratefully acknowledge discussions with Richard Sear. 
MGN acknowledges financial support from EU contract 
ERBFMBICT982949.

%
%

\end{document}